\documentclass[a4paper,fleqn]{cas-sc}

\usepackage[numbers]{natbib}
\usepackage{booktabs}  
\usepackage{multirow}  
\usepackage{caption}   
\usepackage{amsmath}
\usepackage{lineno}
\usepackage{array}
\usepackage{makecell}  
\usepackage{tabularx}
\usepackage{tabulary}

\def\tsc#1{\csdef{#1}{\textsc{\lowercase{#1}}\xspace}}
\tsc{WGM}
\tsc{QE}
\tsc{EP}
\tsc{PMS}
\tsc{BEC}
\tsc{DE}

\begin{document}

\let\WriteBookmarks\relax
\def\floatpagepagefraction{1}
\def\textpagefraction{.001}

\shorttitle{Fast SSVEP Detection Using a Calibration-Free EEG Decoding Framework}

\shortauthors{C. Wang, J. Li et~al.}

\title [mode = title]{Fast SSVEP Detection Using a Calibration-Free EEG Decoding Framework}       




%

\author[1]{Chenlong Wang}[]


\fnmark[1]

\ead{wcl23@mails.tsinghua.edu.cn}



\affiliation[1]{organization={Shenzhen International Graduate School, Tsinghua University},
    city={Shenzhen},
    postcode={518055}, 
    country={China}}


\author[1]{Jiaao Li}[]

\fnmark[1]
\ead{lja23@mails.tsinghua.edu.cn}

\author[4]{Shuailei Zhang}[%
   ]
\ead{shuailei.zhang@ntu.edu.sg}

\author[1]{Wenbo Ding}[]
\ead{ding.wenbo@sz.tsinghua.edu.cn}

\author[1,2,3]{Xinlei Chen}[%
   ]
\cormark[1]
\ead{chen.xinlei@sz.tsinghua.edu.cn}


\affiliation[2]{organization={Pengcheng Laboratory},
    city={Shenzhen},
    postcode={518055}, 
    country={China}}

\affiliation[3]{organization={RISC-V International Open Source Laboratory},
    city={Shenzhen},
    postcode={518055}, 
    country={China}}

\affiliation[4]{organization={College of Computing and Data Science, Nanyang Technological University},
    postcode={639798}, 
    country={Singapore}}

\cortext[cor1]{Principal corresponding author}

\fntext[fn1]{Chenlong Wang and Jiaao Li made equal contributions as first co-authors.}


\begin{abstract}
Steady-State Visual Evoked Potential is a brain response to visual stimuli flickering at constant frequencies. It is commonly used in brain-computer interfaces for direct brain-device communication due to their simplicity, minimal training data, and high information transfer rate. Traditional methods suffer from poor performance due to reliance on prior knowledge, while deep learning achieves higher accuracy but requires substantial high-quality training data for precise signal decoding. In this paper, we propose a calibration-free EEG signal decoding framework for fast SSVEP detection. Our framework integrates Inter-Trial Remixing \& Context-Aware Distribution Alignment data augmentation for EEG signals and employs a compact architecture of small fully connected layers, effectively addressing the challenge of limited EEG data availability. Additionally, we propose an Adaptive Spectrum Denoise Module that operates in the frequency domain based on global features, requiring only linear complexity to reduce noise in EEG data and improve data quality.
\textcolor{blue}{For calibration-free classification experiments on short EEG signals from three public datasets, our framework demonstrates statistically significant accuracy advantages(p<0.05) over existing methods in the majority of cases, while requiring at least 52.7\% fewer parameters and 29.9\% less inference time. By eliminating the need for user-specific calibration, this advancement significantly enhances the usability of BCI systems, accelerating their commercialization and widespread adoption in real-world applications.}


\end{abstract}




\maketitle

\section{Introduction}

Brain-computer interface (BCI) is an innovative neurotechnological system that establishes direct communication between the human brain and external devices. This system decodes neural signals from the brain and converts them into commands to control computers, robotic arms, and other assistive technologies(\cite{daly2008brain,yao2020bacomics, abiri2019comprehensive}). In a typical BCI system, subjects use signal acquisition devices to collect physiological signals, which are then transmitted to a computing device for signal processing. The processed signals are converted into control commands to manipulate external mechanical devices(\cite{narui_wheelchair,sakkalis2022augmented,sunying_cross}). During this process, feedback is provided to the subjects through visual, auditory, and tactile modalities, enabling them to flexibly alter the control commands and adapt to the external environment(\cite{yu2015enhanced,na2021wearable}). BCI enables individuals with severe motor disabilities to interact with their environment more effectively. BCI not only enhances mobility and communication for disabled individuals but also holds promise for rehabilitation and cognitive enhancement(\cite{lin2021cnn,mane2020bci,carelli2017brain}). As research in this field continues to advance, BCI is expected to offer even broader applications, further integrating into everyday life and significantly improving the quality of life for users.

The Steady-State Visual Evoked Potential (SSVEP) paradigm is extensively used in non-invasive BCI systems(\cite{zhang2012multiple,kwak2014toward,vinoj2019brain,lim2017emergency}). It relies on the brain's response to visual stimuli flickering at specific frequencies, causing the visual cortex to generate electrical signals at corresponding frequencies(\cite{regan1989evoked}). This synchronization facilitates precise decoding of the user's attention and intentions in BCI applications. SSVEP-based BCIs are particularly advantageous due to their robust signals, minimal training requirements, and high information transfer rates (ITR) for efficient interactions.

Numerous algorithms have been developed to decode SSVEP signals and perform classification tasks. One widely used traditional approach is Canonical Correlation Analysis (CCA), which maximizes the correlation between recorded EEG signals and predefined reference signals corresponding to specific stimulus frequencies. CCA-based methods (\cite{CCA,FBCCA}) identify the frequency with the highest correlation to classify the SSVEP signal. Although CCA-based methods require no training and operate quickly, their accuracy decreases with shorter signal length, making them unsuitable for BCIs requiring fast response times. Another category of methods is based on trainable spatial filters that classifies SSVEP signals by calculating correlation coefficients. The Task-related Component Analysis (TRCA) designs spatial filters to enhance the similarity between SSVEP signals, thereby improving the signal quality(\cite{nakanishi_TRCA}). Correlated Component Analysis (CORCA) aims to reduce background noise by maximizing the correlation between signals through spatial filtering(\cite{zhang2018correlated}). Task-discriminant Component Analysis (TDCA) utilizes multi-class linear discriminant analysis to create spatiotemporal filters, which facilitates more effective classification of different signal types(\cite{liu2021improving}).



\textcolor{blue}{The advancement of artificial intelligence and increased computational resources have led to the widespread adoption of deep learning-based models across various tasks. The Vision Transformer (ViT) method is used for efficient image classification at scale(\cite{dosovitskiy2020image}). Depth estimation models like Depth Anything have significantly advanced perception capabilities in complex environments(\cite{yang2024depth,luo2024eventtracker}). Weak supervision techniques are employed for robust speech recognition systems that perform effectively across diverse scenarios(\cite{radford2023robust}). Reinforcement learning helps agents continually adjust their strategies in dynamic environments, adapting to environmental changes(\cite{chen2022deliversense,matsuo2022deep,ren2023scheduling}). Large language models are being used to develop more comprehensive AI systems(\cite{zhao2025urbanvideo, touvron2023llama, gao2024embodiedcity}). AI models also enable automatic control and intelligent sensing in robotics(\cite{wang2025ultra,zhu2021deep}), even facilitating coordinated control of multiple robots to complete complex tasks(\cite{chen2015drunkwalk,li2021message,wang2024transformloc}). These technological advancements have also driven significant developments in EEG decoding.}
EEGNet is a compact convolutional neural network tailored for EEG-based BCI, demonstrating significant performance across four distinct BCI paradigms. By utilizing trainable deep convolutional layers instead of traditional frequency and spatial filters, EEGNet enhances its ability to automatically extract relevant features from raw EEG data(\cite{lawhern2018eegnet}). Building on this work, another study adapts EEGNet for asynchronous SSVEP-BCI tasks, marking the first application of convolutional neural networks in this specific context(\cite{waytowich2018compact}). Further advancements are made by Guney, who introduces an innovative neural network architecture for processing multi-channel SSVEP signals. The model uses two steps: pre-training to build a foundation, and fine-tuning to adapt to each subject's SSVEP signals to improve accuracy(\cite{guney2021deep}). Additionally, TRCA-Net combines spatial filtering algorithms with deep learning techniques to improve EEG signal decoding(\cite{deng2023trca}). \textcolor{blue}{DDGCNN introduces layer-wise dynamic graphs to address the over-smoothing problem in Graph Convolutional Networks (GCNs). It implements dense connections to mitigate gradient vanishing issues and enhances feature extraction through graph dynamic fusion(\cite{zhang2024dynamic}). TFF employs a transformer-based architecture to integrate time-frequency features, supporting both SSVEP and RSVP decoding(\cite{TFFformer}). Song proposes EEGConformer, a hybrid network combining CNN and transformer architectures, which comprehensively models EEG signal representations(\cite{song2022eegconformer}).
EEG-Deformer incorporates a Hierarchical Coarse-to-Fine Transformer (HCT) module with Fine-grained Temporal Learning (FTL) branches integrated into Transformers. This architecture effectively distinguishes temporal patterns from coarse to fine scales. It also includes a Dense Information Purification (DIP) module that utilizes multi-level temporal information purification to enhance decoding accuracy(\cite{ding2024eegdeformer}). This approach addresses the limitation of CNN-Transformer 
hybrid networks in effectively capturing coarse-to-fine temporal dynamics in EEG signals.}

Existing deep learning based methods face multiple challenges in EEG decoding. Current algorithms typically focus on calibration-based SSVEP signal classification. When encountering new subjects, these methods require collecting new EEG signals to calibrate model parameters and improve classification accuracy. However, collecting calibration data is time-consuming, primarily because EEG signal quality can vary greatly between individuals(\cite{hsu2015evaluate}).
\textcolor{blue}{For example, while TRCA can achieve high classification accuracy, when calibrating for new subjects, it requires collecting EEG signals from 5 trials for each SSVEP signal category. In tasks with numerous classes (such as 40 classes), this means collecting hundreds of trials, which could take an hour or even longer in practice.} This variability necessitates considerable effort in selecting suitable participants who can generate reliable EEG signals. Moreover, in non-invasive brain-computer interface systems, patch-type wet electrodes, while providing the best EEG signal quality, are difficult to set up and use. The lengthy signal acquisition process often induces negative emotions in subjects and can adversely affect EEG data quality.

Calibration-free EEG decoding algorithms offer a potential solution, as they can be applied to new subjects immediately after initial training without additional calibration. However, current calibration-free approaches face significant limitations in achieving high decoding accuracy. EEG signals exhibit high variability, with notable feature differences across subjects and experimental conditions. This variability challenges both traditional and deep learning algorithms, as models struggle to capture subtle signal features from limited data. Additionally, EEG signals contain various types of complex noise, such as motion artifacts and electromagnetic interference, that traditional denoising methods fail to address effectively. Training with noisy data increases the risk of overfitting and impacts model performance.

Furthermore, most existing EEG decoding algorithms are designed for EEG signals with acquisition times of 1 second or longer. Although longer EEG signals contain more features and can achieve higher decoding accuracy, they inevitably increase the response time of BCI systems. While using shorter EEG signals can improve BCI system response times, which further compounds the challenges of insufficient training data and excessive noise. \textcolor{blue}{Calibration-based methods struggle to process short EEG signals due to their limited signal feature extraction capabilities. For example, TRCA relies on maximizing cross-trial covariance, but shorter signals reduce the number of time points, diminishing the statistical reliability of covariance estimates. This leads to suboptimal spatial filter optimization and weakens the ability to suppress background noise. EEGNet primarily relies on temporal convolution and depends on implicit network learning to suppress frequency-domain noise. However, short signals have low spectral resolution, making it difficult to distinguish between noise and valid components in adjacent frequency bands. Additionally, the model lacks explicit constraints on frequency band weights, potentially retaining frequency bands with high noise content. Furthermore, the segmentation of short signals may reduce the number of tokens, limiting TFF's self-attention mechanism in capturing long-range temporal dependencies. These limitations restrict the deployment of deep learning models in calibration-free, real-time BCI systems}, which hinders the advancement of BCIs in human-computer interaction systems.

\begin{figure*}[t]
    \centering
    \includegraphics[width=1.0\linewidth]{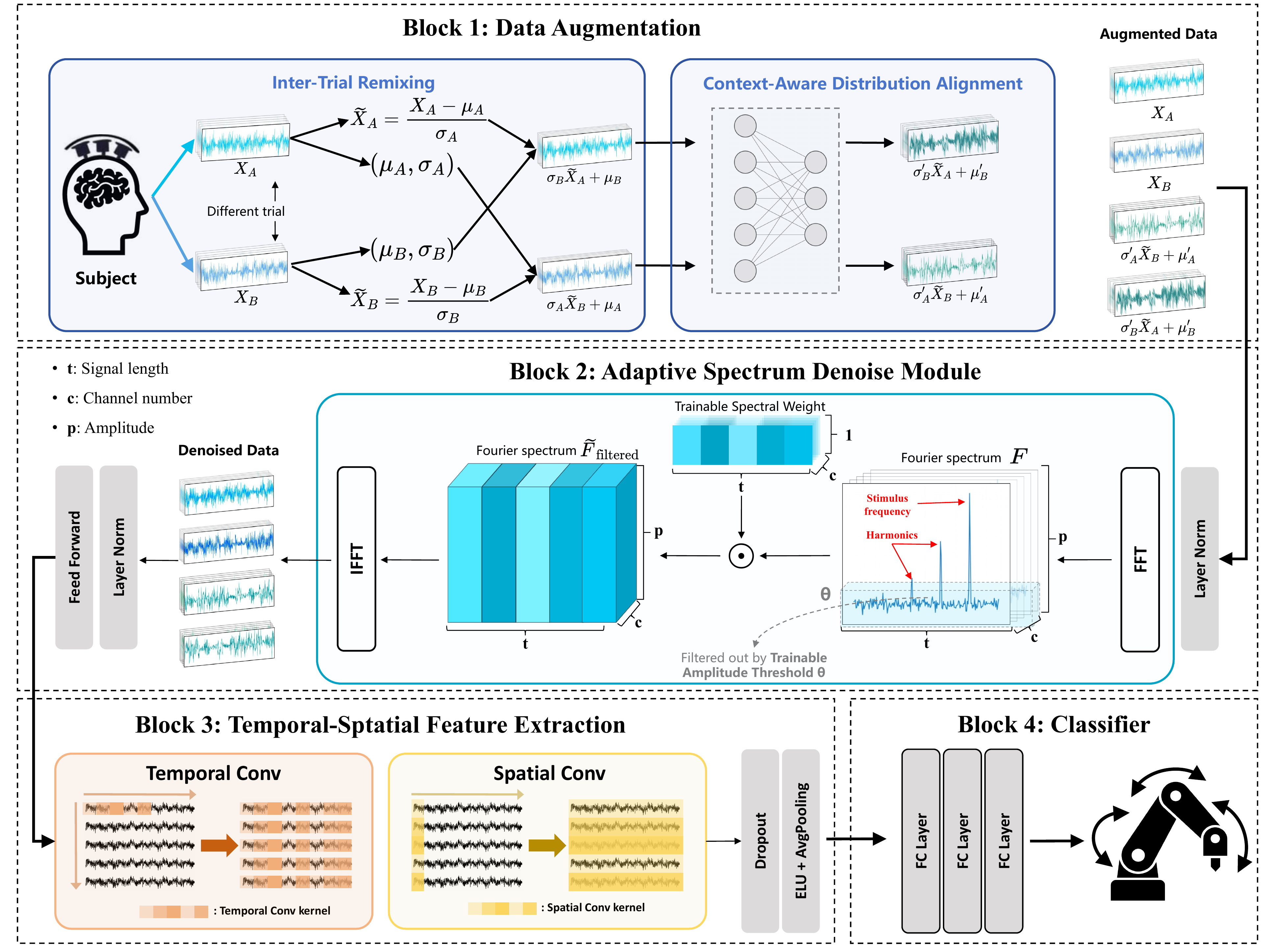}
    \caption{The structural diagram of our proposed calibration-free EEG decoding framework consists of four sequentially connected blocks: Inter-Trial Remixing \& Context-Aware Distribution Alignment Data Augmentation, Adaptive Spectrum Denoise Module, Temporal-Spatial Feature Extraction, and Classifier. Each part contributes to the final predicted classification result.
    }
    \label{fig:whole_network}
\end{figure*}

In this paper, we propose a calibration-free EEG signal decoding framework for fast SSVEP detection. As illustrated in Figure \ref{fig:whole_network}, our framework is composed of four sequentially connected components. The first component is a data augmentation block inspired by image augmentation techniques. It helps the model acquire vital task-relevant content from EEG signals by exchanging data distribution information between trials, and fine-tunes the data distribution to help the model adapt to the exchanged data. The second component is the Adaptive Spectrum-Denoise Module(ASDM), which utilizes spectral analysis to adaptively filter noise based on global information. This noise removal enhances data representation and facilitates feature extraction.  The third component is a feature extraction module that employs convolutional layers of varying shapes along different dimensions as temporal and spatial filters to extract local EEG features. The fourth component is a simple classifier composed of fully connected layers, which outputs the classification results.  

\textcolor{blue}{We achieve calibration-free by addressing two core challenges in EEG signal processing: 1) the high variability of EEG signals, and 2) the substantial noise present in EEG signals. The data augmentation module decouples  trial-specific content from vital task-relevant content by remixing feature map statistics between different trials. This helps the model retain task-relevant information during training and mitigates the negative impact of high EEG signal variability. The ASDM removes noise in the frequency domain using a trainable amplitude threshold and trainable spectral weighting, thus protecting the model from the influence of noise.}

The contributions are summarized as follows:
\begin{itemize}
\item We propose a novel calibration-free EEG decoding framework for fast SSVEP detection. Our framework allows direct application to new subjects without additional data collection for model recalibration. By utilizing short EEG signal, the framework achieves superior SSVEP detection accuracy while maintaining faster response times. Our work significantly improves the usability and accessibility of non-invasive BCI systems, laying the foundation for more practical and user-friendly human-computer interaction technologies.


\item We propose a novel data augmentation method consisting of two steps: Inter-Trial Remixing and Context-Aware Distribution Alignment. The Inter-Trial Remixing step exchanges mean and standard deviation values between different EEG signal trials, helping the model extract vital task-relevant content and enhance robustness during training. The Context-Aware Distribution Alignment step employs a simple attention mechanism to rescale EEG data distributions, effectively aligning statistical characteristics of EEG signals and improving model generalization across different experimental conditions. This method significantly enhances model generalization and robustness, improving model performance with limited EEG data.

\item We propose an Adaptive Spectrum Denoising Module that leverages frequency domain information for effective noise suppression in EEG signals. The module consists of two key components: trainable amplitude thresholds and \textcolor{blue}{trainable spectral weighting}. First, the trainable amplitude thresholds filter out low-amplitude frequency components while preserving key frequencies related to SSVEP signals. Then, \textcolor{blue}{trainable spectral weighting} dynamically \textcolor{blue}{adjusts} the remaining frequency components, further suppressing residual noise and highlighting important frequency features. Finally, an inverse Fourier transform (IFFT) reconstructs the processed frequency-domain signal back to the time domain, generating clearer and refined signal representations.

\item We conducted extensive experimental evaluations using three public datasets to validate the effectiveness of our proposed framework in calibration-free, short EEG signal decoding. Our experiments compared the framework against both traditional and deep learning methods. Experimental results demonstrate that our framework significantly outperforms these methods in classification accuracy, model size, and inference speed. The framework exhibits superior stability and robustness, particularly when processing short-duration signals and in high-noise environments. We also conducted ablation studies to verify the roles of the data augmentation module and adaptive spectrum denoising module in enhancing framework performance. 
\end{itemize}

\section{Method}

\subsection{Preliminaries}
\label{sec:preliminaries}

The \textbf{Fast Fourier Transform (FFT)} is a highly efficient algorithm for computing the Discrete Fourier Transform (DFT), which is a fundamental tool in signal processing for analyzing the frequency content of signals. By reducing the computational complexity of the DFT from $O(N^2)$ to $O(N \log N)$, the FFT makes it feasible to process large datasets and real-time signals. This efficiency is particularly beneficial in applications such as digital signal processing, image processing, and communications, where fast and accurate frequency analysis is essential.

The DFT of a discrete time-domain signal $x[n]$ of length $N$ is given by:

\begin{equation}
X[k] = \sum_{n=0}^{N-1} x[n] e^{-j \frac{2\pi}{N} kn}, \quad k = 0, 1, \ldots, N-1
\end{equation}

The FFT exploits the symmetry and periodicity properties of the DFT to speed up this computation. The most common implementation of the FFT is the Cooley-Tukey algorithm, which recursively breaks down a DFT of size $N$ into smaller DFTs, significantly reducing the number of operations required.

The \textbf{Inverse Fast Fourier Transform (IFFT)} is an algorithm for efficiently computing the inverse of the DFT, transforming a frequency-domain signal $X[k]$ back into its time-domain representation. The IFFT is computed using the same principles as the FFT, but with the input and output interchanged, and the twiddle factors conjugated. The inverse DFT is given by:

\begin{equation}
x[n] = \frac{1}{N} \sum_{k=0}^{N-1} X[k] e^{j \frac{2\pi}{N} kn}, \quad n = 0, 1, \ldots, N-1
\end{equation}

Both the FFT and IFFT are essential in a wide range of applications, including filtering, spectral analysis, image compression, and solving partial differential equations, where frequency-domain representations can simplify complex operations.

\subsection{Data augmentation}

\textcolor{blue}{In this section, we introduce a novel data augmentation method that addresses the challenge of insufficient training data, significantly improving the generalizability and robustness of the model. This achieves better performance in handling short EEG signals and supports calibration-free application scenarios. Our proposed method consists of two steps: Inter-Trial Remixing and Context-Aware Distribution Alignment.}


\subsubsection{Inter-Trial Remixing}
In this article, vital task-relevant content refers to features crucial to specific EEG decoding tasks, whereas trial-specific content is influenced by external conditions. The purpose of Inter-Trial Remixing is to help the model extract vital task-relevant content from raw signals during training while appropriately disregarding trial-specific content.  

Specifically, a sample of the EEG signal can be represented as a tensor \( \mathbf{X} \in \mathbb{R}^{C \times T} \). \( C \) represents the number of channels, indicating the number of electrodes on the EEG cap in EEG decoding tasks. \( T \) represents the number of sampling points, i.e., the signal length. Assume there are two samples of EEG signals from different trials of the same subject, denoted as \( \mathbf{X}_A \) and \( \mathbf{X}_B \), respectively. As shown in Figure \ref{fig:whole_network} Block 1, we normalize \( \mathbf{X}_A \) and \( \mathbf{X}_B \), generating \( \widetilde{\mathbf{X}}_A \) and \( \widetilde{\mathbf{X}}_B \). Subsequently, we multiply \( \widetilde{\mathbf{X}}_A \) by the standard deviation of \( \mathbf{X}_B \) and add the mean of \( \mathbf{X}_B \), and perform the corresponding operation for \( \widetilde{\mathbf{X}}_B \). This results in \( \sigma_B \widetilde{\mathbf{X}}_A + \mu_B \) and \( \sigma_A \widetilde{\mathbf{X}}_B + \mu_A \). 
\textcolor{blue}{This step exchanges the mean and standard deviation between different trials, preserving signal trends and periodic nature while altering  trial-specific content. It is important to note that Inter-Trial Remixing is used only during training to expand data distribution. For testing, we narrow the distribution to promote trial-specific content uniformity through the Context-Aware Distribution Alignment module.}


\subsubsection{Context-Aware Distribution Alignment}

Due to the high variability and instability of EEG signals, signals under different experimental conditions exhibit markedly different distributions. This presents a significant challenge for EEG decoding models. As shown in Figure \ref{fig:whole_network} Block 1, in the Context-Aware Distribution Alignment step, we apply several fully connected layers to establish a simple attention mechanism for \( \sigma \tilde{\mathbf{X}} + \mu \). This mechanism calibrates the mean \( \mu \) and the standard deviation \( \sigma \), expressed as \( \mu' = f(\mu, \sigma)\mu \) and \( \sigma' = g(\mu, \sigma)\sigma \), where \( f \) and \( g \) represent the linear transforms with shape of (Channel, Channel). The calibrated features can thus be represented as: \( \sigma'\widetilde{X} + \mu' \)

The \textcolor{blue}{Context-Aware Distribution Alignment} step employs a computationally efficient approach to \textcolor{blue}{modulate} the statistical properties of the EEG signals. \textcolor{blue}{Unlike traditional normalization techniques that apply fixed transformations, this module adapts its fitting operation based on the statistical context of the input EEG signal. Specifically, it learns to modulate the mean and variance of each input EEG dynamically through fully connected layers, enabling the model to reconstruct realistic and consistent distributions(\cite{crossnorm_selfnorm}). This context-awareness ensures that the remixed signals are rescaled in a way that preserves vital task-relevant content while suppressing trial-specific noise in EEG.}


\subsection{Adaptive Spectrum Denoise Module}

\textcolor{blue}{Computing the Fourier spectrum utilizes global EEG signal information, but lacks adaptive weight allocation capabilities compared to transformers. Our proposed Adaptive Spectrum-Denoise Module leverages the EEG signal spectrum for adaptive denoising from a global perspective. As shown in Figure \ref{fig:whole_network} Block 2, the module consists of layer normalization, trainable amplitude filtering, trainable spectral weighting, and feedforward layer. We apply trainable amplitude filtering to remove most noise from the Fourier spectrum, then use trainable spectral weighting to suppress residual noise by adaptively weighting frequency components. Finally, we apply IFFT to reconstruct the signal, effectively removing noise and enhancing representations.}

\subsubsection{Trainable amplitude thresholds}
EEG signals generated by the SSVEP phenomenon exhibit distinct characteristics in the frequency domain, particularly that the amplitude of the frequency component corresponding to the stimulus frequency is significantly larger than other frequency components. This is typically observed at the stimulus frequency and its harmonics(\cite{muller2005steady}). For example, if a subject is exposed to a 12 Hz flickering light, the EEG will show a peak at 12 Hz and smaller peaks at harmonics like 24 Hz and 36 Hz, while other unrelated frequencies will have significantly lower amplitudes, reflecting the brain's selective response to the stimulus. Therefore, filtering the EEG signal based on amplitude can effectively extract significant features. We establish a threshold in the frequency domain, where frequency components with amplitudes greater than this threshold are retained, while those below the threshold are filtered out. To enhance the model's adaptability, we set this threshold as a trainable parameter. As section \ref{sec:preliminaries} mentioned, by utilizing FFT, we can obtain the Fourier spectrum of the EEG signal $F$, and its power spectrum as $P$ ,which can be expressed by:

\begin{align}
F[k] &= \sum_{n=0}^{N-1} x[n] e^{-j \frac{2\pi}{N} kn}, \quad k = 0, 1, \ldots, N-1
\end{align}

\begin{align}
\textcolor{blue}{P[k]=\frac{1}{N}\cdot|F[k]|^2,\quad k=0,1,\dots,N-1}
\end{align}
\noindent \textcolor{blue}{where $F[k]$ represents the $k$-th coefficient of the Discrete Fourier Transform, $P[k]$ denotes the power spectrum at frequency index $k$, $x[n]$ is the amplitude of the EEG signal at time point $n$, $N$ is the total number of time points in the signal, $e^{-j \frac{2\pi}{N} kn}$ is the complex exponential term with $j = \sqrt{-1}$ being the imaginary unit, $k$ represents the frequency index ranging from $0$ to $N-1$, $n$ represents the time index ranging from $0$ to $N-1$, and $|\cdot|^2$ denotes the squared magnitude of a complex number.  }

\textcolor{blue}{To enhance the model's adaptability, we set $\theta$ as a trainable parameter, initializing it with a random value between (0,1).} We then take the median of $P$, and utilize it to normalize $P$ to obtain $P_m$. This can be expressed by:

\begin{align}
&P_m = \frac{P}{\text{Median}(P)}
\end{align}

then the filtering process can be expressed by:
\begin{equation}
F_{filtered}[k] = F[k]*(P_m[k]>\theta ) 
\end{equation}
\textcolor{blue}{where} $P_m[k]>\theta$ is a binary mask, which equals 1 if the inequality holds, and 0 otherwise.  

\begin{figure*}[t]
    \centering
    \includegraphics[width=1.0\linewidth]{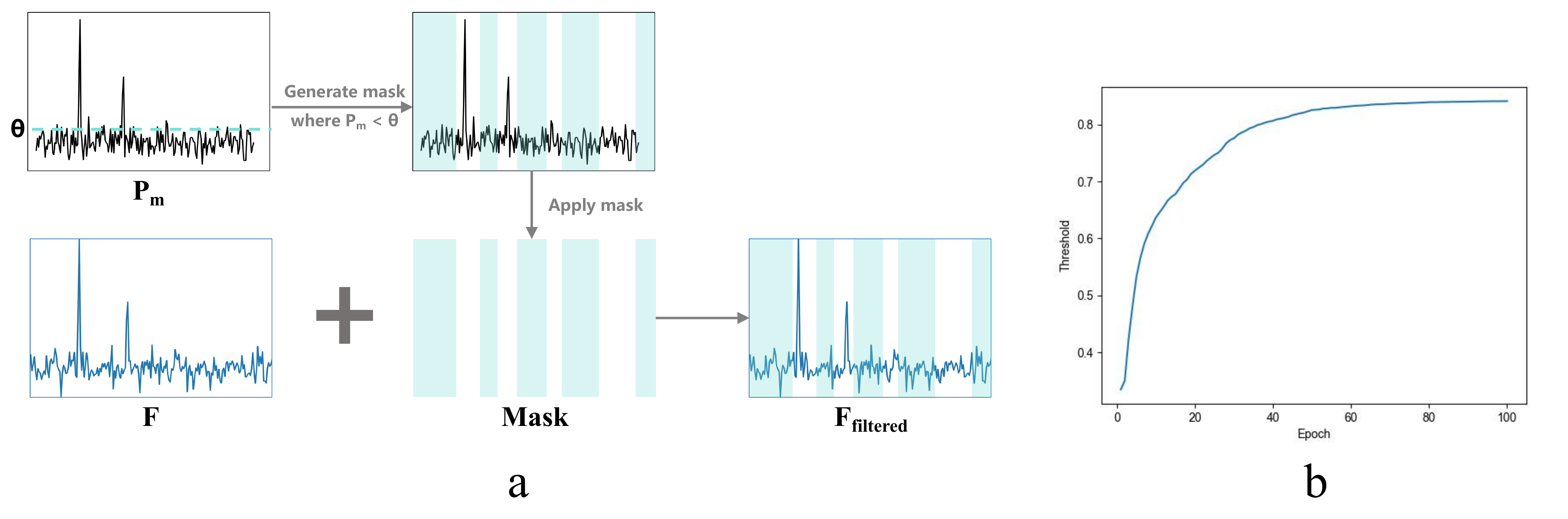}
    \caption{(a) Flow diagram of the Trainable Amplitude Threshold Filter. (b) The learning curve of the threshold parameter $\theta$ during the training process, where the x-axis represents training epochs and the y-axis shows the threshold values.
    }
    \vspace{-15pt}
    \label{fig:threshold}
\end{figure*}

\textcolor{blue}{
Figure \ref{fig:threshold}.a demonstrates the process of applying a trainable amplitude threshold. In this example, when the model is training and $\theta$ reaches 0.9 during optimization, it generates a mask for positions in the normalized power spectrum $P_m$ with values less than 0.9. The corresponding positions in $F$ are set to zero, producing $F_{filtered}$. Figure \ref{fig:threshold}.b shows the learning curve of $\theta$, which illustrates how $\theta$ converges to an optimal value during training. This optimal threshold effectively filters out most frequency-domain noise while preserving the frequency components relevant to SSVEP tasks.}
\subsubsection{Trainable Spectral Weighting}
After the initial amplitude filtering, we obtain the Fourier spectrum $F_{\text{filtered}}$ containing significant features. However, due to individual differences, some subjects have harmonic components with very low amplitude and significant noise from other frequency components, which can prevent the amplitude filter from eliminating all noise. Additionally, the most informative frequency components of SSVEP signals are concentrated around the fundamental and harmonic frequencies of the stimulus, necessitating a frequency-specific weight allocation module to extract more effective SSVEP features. Therefore, we propose Trainable Spectral Weighting to enhance the module's frequency adaptability, allowing for personalized frequency selection while reducing noise impact.  
Specifically, for $F_{\text{filtered}}$, we introduce a trainable weight $M_f$, which adaptively assigns weights to different frequency components along the frequency dimension of $F_{\text{filtered}}$. This can be expressed as:


\begin{align}
&\widetilde{F}_{\text{filtered}}  = F_{\text{filtered}} \odot M_{f}
\end{align}

Here, $\odot$ denotes element-wise multiplication. This process adaptively assigns trainable weights to each frequency component in the frequency domain, allowing the model to emphasize important frequency bands while suppressing irrelevant or noisy components.

Finally, we need to convert the frequency-domain features $\widetilde{F}_{\text{filtered}}$ into the time domain to serve as time-domain tokens for input into the layer normalization. We achieve this using the inverse fast Fourier transform (IFFT)

\begin{align}
x[n] &= \frac{1}{N} \sum_{k=0}^{N-1} \widetilde{F}_{\text{filtered}}[k] e^{j \frac{2\pi}{N} kn}, \quad n = 0, 1, \ldots, N-1
\end{align}

The Adaptive Spectrum Denoise Module improves EEG signal processing using trainable spectral analysis. It applies trainable amplitude thresholds to filter out noise and \textcolor{blue}{trainable spectral weighting} to highlight critical frequency components. \textcolor{blue}{Among existing methods, TFF only uses cross-attention to merge time-frequency domain features without specifically designing for SSVEP's spectral characteristics. Merging temporal features actually makes the model performance unstable (see Section \ref{sec:comparisons}). In contrast, ASDM employs trainable amplitude thresholds to directly filter out non-target frequencies. It also combines trainable spectral weighting to adaptively allocate importance along the frequency axis for fundamental stimuli frequencies, their harmonics, and other frequency bands, effectively removing residual noise. While TRCA maximizes spatial covariance across multiple trials from the same subject to improve SNR, it cannot dynamically adapt to individual differences. ASDM addresses this limitation through trainable spectral weight matrices that dynamically adjust frequency domain features, enhancing cross-subject generalization capabilities. While TRCA maximizes spatial covariance across multiple trials from the same subject to improve SNR, it cannot dynamically adapt to individual differences. ASDM addresses this limitation through trainable spectral weight matrices that dynamically adjust frequency domain features, enhancing cross-subject generalization capabilities.}  

The filtered signals are then reconstructed via the inverse Fourier transform, producing cleaner and more refined signal representations.

\subsection{Feature Extraction}
We use two 1D convolutions moving along different dimensions as adaptive filters to extract EEG features, referred to as Temporal Conv and Spatial Conv respectively(\cite{thomas2017deep, lawhern2018eegnet,song2022eegconformer}). As illustrated in Figure \ref{fig:whole_network} Block 3, the convolutional kernel of the temporal convolution moves along the temporal dimension with a shape of $(1, d)$ and a stride of $(1, 1)$, where $d$ is the length of the kernel in the temporal dimension, indicating that $d$ sampling points are processed at a time. After the temporal convolution kernel finishes extracting features from one channel, it moves on to the next channel. The shape of the spatial convolution kernel is $(c, 1)$ with a stride of $(1, 1)$, where $c$ is the number of channels, i.e., the number of electrodes. The spatial convolution processes data from all electrode channels simultaneously to learn the representation of the interactions between different electrode channels. After temporal and spatial filtering, we apply the Exponential Linear Unit (ELU) as the activation function to introduce nonlinearity into the model, which is defined by the following equation:
\begin{equation}
f_{ELU}(x) = \begin{cases} 
      x, & \text{if } x > 0 \\
      a(e^x - 1), & \text{if } x \leqslant 0
   \end{cases}
\end{equation}
Due to the normalization process and the high variability of EEG data, a significant portion of the EEG features contains negative values. As a result, using the rectified linear unit (ReLU) as the activation function would lead to a loss of feature information. By allowing negative values, ELUs enable the propagation of information and gradients even when the input is negative, thus maintaining the flow of gradients throughout the network.
Additionally, we use an average pooling layer to smooth the temporal features and reduce the model's complexity.  


The feature extraction block utilizes the local receptive field of convolutional kernels to extract features from EEG signals across different dimensions. It focuses on the local spatiotemporal patterns of the signal, which works in synergy with the Adaptive Spectrum-Denoise Module.

\subsection{Classifier}
\textcolor{blue}{Finally, following the established practice in mainstream EEG classification models(\cite{lawhern2018eegnet, song2022eegconformer})}, we use three cascaded fully connected layers as the classifier to aggregate and compress features generated by the feature extraction block, apply the softmax function to generate a probability vector, and then obtain the classification results. During the training process, we use cross-entropy as the loss function, which can be expressed by the following equation:

\begin{align}
L_{cross\_entropy} = - \frac{1}{N} \sum_{n=1}^{N} \sum_{r=1}^{R} y_{n,r} \log \hat{y}_{n,r}
\end{align}
where $N$ is the number of EEG signal-ground truth pairs in each minibatch, $R$ is the total number of EEG signal categories, and $\hat{y}_{n,r}$ and $y_{n,r}$ represent the probability vector predicted by the framework and the ground truth, respectively.

\section{Experiments}
\subsection{Experiment Settings}
Our experiments used a server equipped with an Intel Xeon Gold 6242R CPU and an NVIDIA RTX A6000 GPU for training and testing. We implemented the framework in Python 3.9.19 with PyTorch 1.13.1. The optimization process utilized the Adam optimizer with $\beta_{1}$ and $\beta_{2}$ values of 0.9 and 0.999, respectively. Before feeding data into the framework, we preprocessed the signals using a Butterworth bandpass filter. In the inference speed evaluation experiments, we tested the framework on 240 EEG signal samples, representing the complete data collected from one subject in the Benchmark dataset. To ensure fair comparison across all methods, we conducted all inference time measurements exclusively on CPU.
\subsection{Dataset}
\subsubsection{Benchmark Dataset}
The Benchmark Dataset (\cite{benchmarkdataset}) for SSVEP-Based BCI is a comprehensive dataset designed for SSVEP research in brain-computer interfaces. It includes 64-channel EEG data from 35 healthy participants performing a cue-guided target selection task with a 40-target BCI speller. Targets are encoded using joint frequency and phase modulation (JFPM) within the 8-15.8 Hz range at 0.2 Hz intervals, with adjacent targets having phase offsets of 0.5$\pi$ for clear differentiation. EEG data are recorded using a 64-electrode system at a 1000 Hz sampling rate, later downsampled to 250 Hz (\cite{nakanishi_TRCA}). This dataset is valuable as it provides synchronized visual stimulation and EEG data for each frequency and phase combination, enabling precise simulation of new stimulation phases. 




When using the Benchmark dataset, we \textcolor{blue}{adopt} a leave-one-subject-out cross-validation approach. Specifically, for each iteration, we \textcolor{blue}{use} EEG signals from one subject as the test set and signals from the remaining 34 subjects as the training set. We \textcolor{blue}{use} the signals collected from eight electrodes as our input data: PO5, PO3, POz, PO4, PO6, O1, Oz, and O2.

\subsubsection{BETA Dataset}
The BETA (Benchmark database Towards BCI Application) dataset (\cite{betadataset}) is a robust resource for advancing SSVEP-based BCI systems. It includes 64-channel EEG data from 70 participants performing a 40-target cued-spelling task using a virtual QWERTY keyboard interface. This setup reflects real-world application settings rather than controlled laboratory conditions, enhancing practical usability and user experience by mimicking conventional input devices. The visual stimuli flicker at frequencies between 8 Hz and 15.8 Hz. Recognizing that frequency resolution does not significantly impact SSVEP classification results (\cite{nakanishi_TRCA}), all epochs \textcolor{blue}{are} downsampled to 250 Hz. 

For the BETA dataset, we also adopt a leave-one-subject-out cross-validation approach. To maintain consistency with the Benchmark dataset, we select signals from the same eight electrodes for our analysis.

\subsubsection{\textcolor{blue}{Nakanishi Dataset}}
\textcolor{blue}{
Nakanishi dataset(\cite{nakanishi2015comparison}) consists of EEG recordings acquired during a visual stimulation experiment where 12 visual target stimuli are displayed on a 27-inch LCD monitor, modulated at frequencies ranging from 9.25 Hz to 14.75 Hz with 0.5 Hz increments. Ten participants with normal or corrected-to-normal vision participate in the experiment, sitting 60 cm from the monitor in a dimly lit room. The BioSemi ActiveTwo EEG system (Biosemi, Inc.) is used to acquire EEG data from eight electrodes in the occipital region. The experiment consists of 15 blocks, with each block containing 12 trials corresponding to the 12 flickering target stimuli. In each trial, subjects are required to gaze at one randomly selected target stimulus for 4 seconds. The EEG signal is initially sampled at 2048 Hz and subsequently downsampled to 256 Hz.
For the Nakanishi dataset, we employ the leave-one-subject-out validation method. Since this dataset only records data from eight electrodes, we utilize all available data for training and testing.
}
\subsection{Compared Methods}
We compared our method with several existing methods on public datasets. CCA(\cite{CCA}) and FBCCA(\cite{FBCCA}) are traditional methods that do not require training and rely solely on prior knowledge for SSVEP classification. TRCA(\cite{nakanishi_TRCA}) is another traditional baseline that requires training and employs spatial filters to reduce environmental noise. In contrast, TFF(\cite{TFFformer}) and EEGConformer(\cite{song2022eegconformer}) are deep learning methods; TFF decodes EEG signals using a two-stream transformer, while EEGConformer combines CNN and attention mechanisms to process both local and global information. Comparative experiments show that our framework demonstrates significant advantages over existing methods in terms of classification accuracy, model size, and inference speed.


\subsection{Classification Performance}
\label{sec:comparisons}
\begin{figure*}[t]
    \centering
    \includegraphics[width=0.9\linewidth]{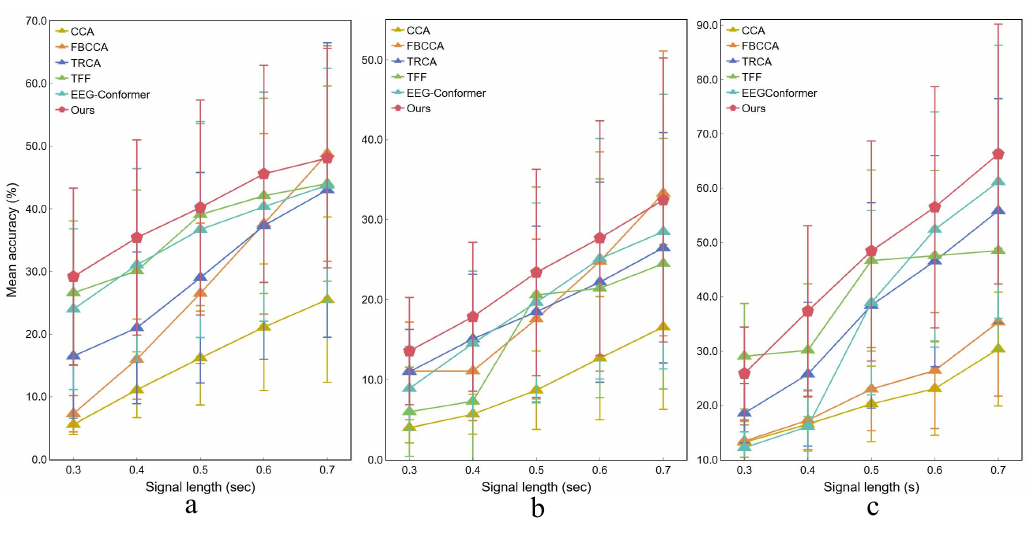}

    \caption{Comparison of Performance Among Different SSVEP Classification Methods. (a) Accuracy and standard deviation on the Benchmark dataset. (b) Accuracy and standard deviation on the BETA dataset. (c) Accuracy and standard deviation on the Nakanishi dataset.
}
    \label{fig:comparison_result}
\end{figure*}

Figure \ref{fig:comparison_result} illustrates the comparative experimental results of our framework against other baselines, including accuracy and variance. We used the Wilcoxon Signed-Rank Test  to assess the statistical significance of the results. Significance levels are denoted by asterisks: * for $p<0.05$, ** for $p<0.01$, and *** for $p<0.001$.  We present the Wilcoxon Signed-Rank Test results for the three datasets in Table \ref{comparison_significance}.

The results on the Benchmark dataset reveal that SSVEP classification accuracy generally increases with signal length, highlighting the challenge of feature extraction from short EEG signals. TRCA outperforms FBCCA and CCA for short signal lengths (<0.6s), but this advantage diminishes or reverses as the sequence lengthens. This indicates that training-free methods based solely on prior knowledge struggle with short EEG signals. Our deep learning-based approach shows significant advantages over traditional methods (TRCA, FBCCA, CCA). This is because they primarily rely on classical statistical feature extraction methods, which cannot fully capture the complex time-frequency characteristics of EEG signals. For instance, at a signal length of 0.3s, our framework achieves 29.2\% $\pm$ 14.12\% accuracy, compared to TRCA (16.5\% $\pm$ 9.9\%), FBCCA (7.3\% $\pm$ 2.9\%), and CCA (5.6\% $\pm$ 1.6\%). These differences of 12.7\%, 21.9\%, and 23.6\% respectively are statistically significant ($p<0.001$, \textcolor{blue}{with confidence intervals of [20.04, 28.13], [18.02, 26.43], and [9.54, 16.42]}). Our framework also outperforms other deep learning-based EEG decoding methods like TFF and EEGConformer across all signal lengths. At 0.4s, our framework achieves (35.4\% $\pm$ 15.6\%) accuracy, significantly surpassing TFF (30.1\% $\pm$ 12.9\%) and EEGConformer (31.0\% $\pm$ 15.4\%), with statistically significant differences of 5.3\% and 4.4\% respectively ($p<0.01$ and $p<0.001$, \textcolor{blue}{with confidence intervals of [0.62, 4.90], [3.99, 6.62]}). At a signal length of 0.7s, FBCCA outperforms all other methods, including our framework. This demonstrates FBCCA's effectiveness in processing longer EEG signals. However, our framework remains the closest competitor to FBCCA, with only a 0.72\% performance difference. This indicates that our framework still performs excellently when processing long EEG signals.   

\begin{figure*}[t]
    \centering
    \includegraphics[width=1.0\linewidth]{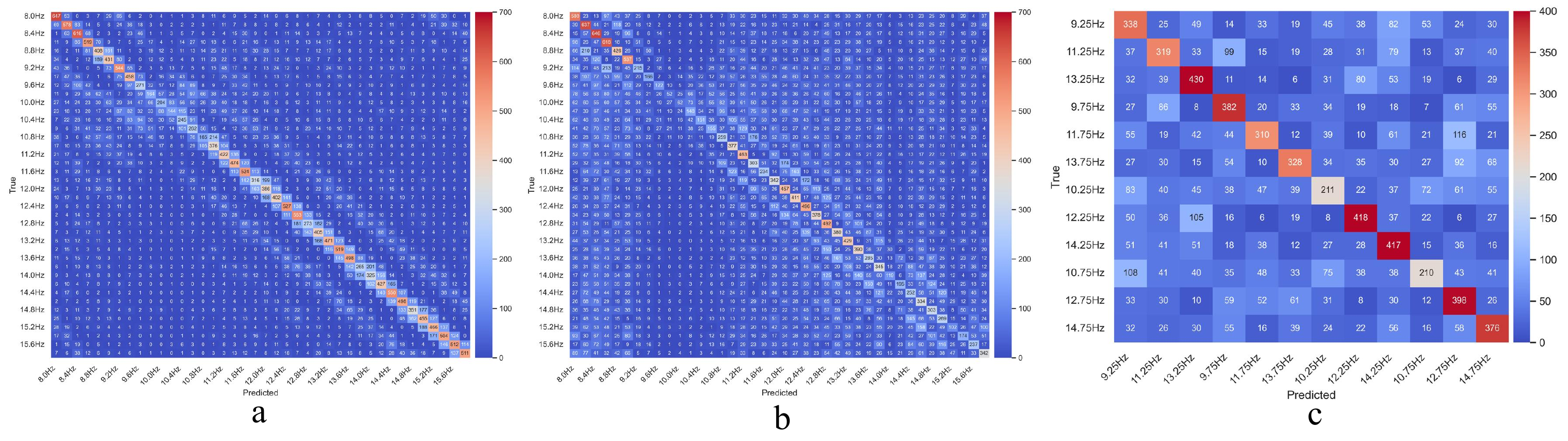}

    \caption{Confusion matrices produced by our framework across different datasets. (a) Benchmark dataset (b) BETA dataset (c) Nakanishi dataset
}
    \label{fig:lwf_confusion_matrix}
\end{figure*}

\begin{figure*}[t]
    \centering
    \includegraphics[width=1.0\linewidth]{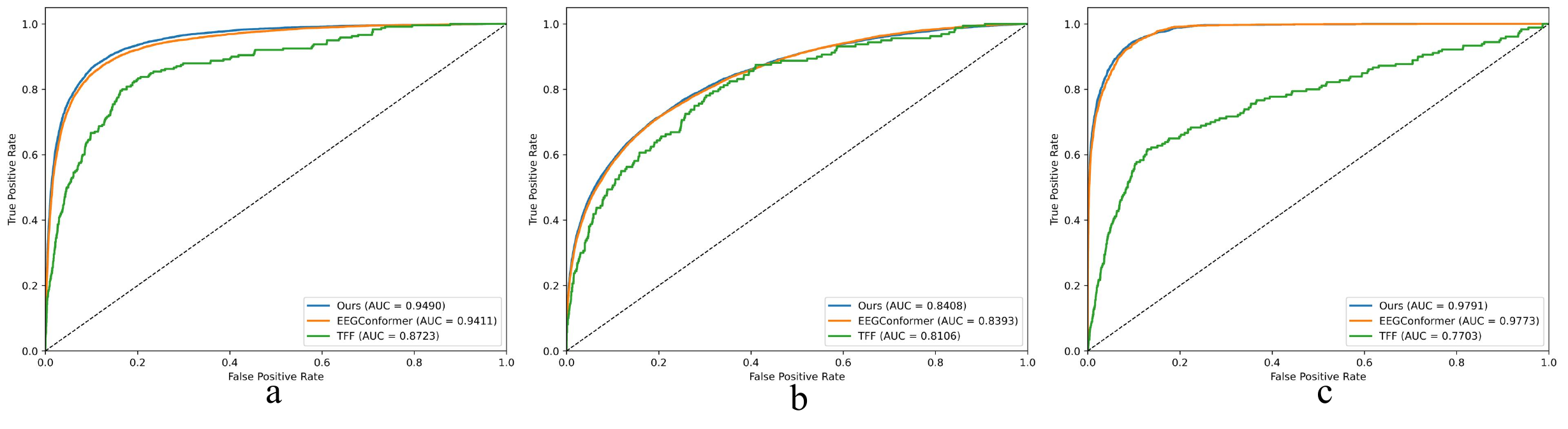}

    \caption{ROC curves of our framework and other baselines on different datasets: (a) Benchmark dataset (b) BETA dataset (c) Nakanishi dataset
}
    \label{fig:roc}
\end{figure*}

\textcolor{blue}{Figure \ref{fig:lwf_confusion_matrix} (a) displays the confusion matrix of our framework on the benchmark dataset. We observe two distinct error patterns in frequency prediction. The first error pattern shows confusion between adjacent frequencies, which is understandable since SSVEP signals evoked by neighboring stimulus frequencies likely have similar representations in the feature space. The second error pattern reveals that stimulus frequencies tend to be confused with frequencies that differ by approximately 1Hz (higher or lower). This occurs because modulation signals between such frequency pairs exhibit higher similarity(\cite{guney2021deep}). Figure \ref{fig:roc} (a) illustrates the ROC curves comparing our framework with TFF and EEGConformer. It's worth noting that CCA, FBCCA, and TRCA generate classification results by comparing correlation coefficients rather than probability vectors through softmax, making ROC curves unavailable for these three methods. The ROC curves demonstrate that our method performs best on the Benchmark dataset, achieving an AUC value of 0.9490, slightly higher than EEGConformer's 0.9411 and significantly better than TFF's 0.8723. Our framework also exhibits higher true positive rates in the low false positive rate region, indicating that our model predicts positive samples with higher confidence.}

Figure \ref{fig:comparison_result} (b) shows the experimental results based on the BETA dataset. Since the BETA dataset was not collected under controlled laboratory conditions, it contains a significant amount of noise. Consequently, all methods perform relatively worse on the BETA dataset compared to the Benchmark dataset. Notably, unlike the Benchmark dataset results, TFF performs significantly worse than traditional methods like TRCA and FBCCA on the BETA dataset. While EEGConformer outperforms TFF, it fails to demonstrate the advantages of deep learning methods in processing short EEG signals. This discrepancy arises from the BETA dataset's higher environmental noise, which directly degrades training data quality and severely impacts deep learning methods' performance. Our method overcomes environmental noise interference through data augmentation, highlighting vital task-relevant content in EEG signals, and employing Adaptive Spectrum Denoise Module for adaptive denoising. For signal lengths $<0.7s$, our framework achieves the highest classification accuracy among all compared methods. Specifically, at a signal length of 0.3s, our method achieves 13.6\% $\pm$ 6.7\% accuracy, compared to TRCA (11.0\% $\pm$ 5.3\%), FBCCA (11.1\% $\pm$ 6.1\%), and CCA (4.0\% $\pm$ 1.9\%). The differences of 2.6\%, 2.5\%, and 9.6\% respectively are statistically significant ($p<0.001$, \textcolor{blue}{with confidence intervals of [6.5, 9.1], [1.1, 4.0], [1.7, 3.8]}). TFF and EEGConformer achieve 6.0\% $\pm$ 5.6\% and 8.9\% $\pm$ 5.3\% accuracy, with differences of 7.6\% and 9.6\% respectively, also statistically significant ($p<0.001$, \textcolor{blue}{with confidence intervals of [6.0, 9.6], [4.0, 5.5]}). Similar conclusions hold up to signal length 0.6s, demonstrating our framework's significant advantage in processing EEG signals with high environmental noise. At signal length 0.7s, FBCCA performs best, but our method remains the closest competitor with only a 0.8\% non-significant difference, similar to the Benchmark dataset results.    

\renewcommand{\arraystretch}{1.2} 
\begin{table}[htbp]
\centering
\fontfamily{ptm}\selectfont
\fontsize{6pt}{6pt}\selectfont
\caption{Wilcoxon Signed-Rank Test results on three public datasets. * indicates significant differences between our framework and baselines: * $p<0.05$, ** $p<0.01$, *** $p<0.001$. - indicates that the difference is not significant.}

\begin{tabularx}{\textwidth}{
  X 
  >{\centering\arraybackslash}X
  >{\centering\arraybackslash}X
  >{\centering\arraybackslash}X
  >{\centering\arraybackslash}X
  >{\centering\arraybackslash}X 
  >{\centering\arraybackslash}X
  >{\centering\arraybackslash}X
  >{\centering\arraybackslash}X
  >{\centering\arraybackslash}X
  >{\centering\arraybackslash}X 
  >{\centering\arraybackslash}X
  >{\centering\arraybackslash}X
  >{\centering\arraybackslash}X
  >{\centering\arraybackslash}X
  >{\centering\arraybackslash}X
}
\toprule
& \multicolumn{15}{c}{\textbf{Signal length(s)}}\\
\cmidrule(lr){2-16}
& \multicolumn{5}{c}{\textbf{Benchmark}} & \multicolumn{5}{c}{\textbf{BETA}} & \multicolumn{5}{c}{\textbf{Nakanishi}}\\
\cmidrule(lr){2-6} \cmidrule(lr){7-11} \cmidrule(lr){12-16}
\textbf{Method} & \textbf{0.3s} & \textbf{0.4s} & \textbf{0.5s} & \textbf{0.6s} & \textbf{0.7s} & \textbf{0.3s} & \textbf{0.4s} & \textbf{0.5s} & \textbf{0.6s} & \textbf{0.7s} & \textbf{0.3s} & \textbf{0.4s} & \textbf{0.5s} & \textbf{0.6s} & \textbf{0.7s}\\
\midrule

CCA & 
$^{{\scriptsize ***}}$\newline{[20.0,28.1]} & 
$^{{\scriptsize ***}}$\newline{[22.0,31.3]} & 
$^{{\scriptsize ***}}$\newline{[21.3,31.9]} & 
$^{{\scriptsize ***}}$\newline{[20.2,31.6]} & 
$^{{\scriptsize ***}}$\newline{[15.4,28.0]} & 
$^{{\scriptsize ***}}$\newline{[6.5,9.1]} & 
$^{{\scriptsize ***}}$\newline{[6.7,10.2]} & 
$^{{\scriptsize ***}}$\newline{[7.0,11.6]} & 
$^{{\scriptsize ***}}$\newline{[5.8,11.1]} & 
$^{{\scriptsize ***}}$\newline{[5.0,11.8]} &
$^{{\scriptsize **}}$\newline{[9.0,16.9]} & 
$^{{\scriptsize **}}$\newline{[14.7,27.5]} & 
$^{{\scriptsize **}}$\newline{[20.6,36.6]} & 
$^{{\scriptsize **}}$\newline{[25.3,42.2]} & 
$^{{\scriptsize **}}$\newline{[28.3,44.1]} \\
\cmidrule{1-16}

FBCCA & 
$^{{\scriptsize ***}}$\newline{[18.0,26.4]} & 
$^{{\scriptsize ***}}$\newline{[14.7,25.0]} & 
$^{{\scriptsize ***}}$\newline{[7.9,20.2]} & 
$^{{\scriptsize *}}$\newline{[1.8,14.7]} & 
$^{{\scriptsize -}}$\newline{[\mbox{-}7.3,6.7]} & 
$^{{\scriptsize ***}}$\newline{[1.1,4.0]} & 
$^{{\scriptsize ***}}$\newline{[5.1,8.6]} & 
$^{{\scriptsize ***}}$\newline{[3.4,8.4]} & 
$^{{\scriptsize *}}$\newline{[0.1,6.0]} & 
$^{{\scriptsize -}}$\newline{[\mbox{-}4.3,3.0]} &
$^{{\scriptsize **}}$\newline{[9.1,16.1]} & 
$^{{\scriptsize **}}$\newline{[13.9,26.7]} & 
$^{{\scriptsize **}}$\newline{[17.4,33.8]} & 
$^{{\scriptsize **}}$\newline{[22.3,38.6]} & 
$^{{\scriptsize **}}$\newline{[23.2,39.6]} \\
\cmidrule{1-16}

TRCA & 
$^{{\scriptsize ***}}$\newline{[9.5,16.4]} & 
$^{{\scriptsize ***}}$\newline{[10.5,18.9]} & 
$^{{\scriptsize ***}}$\newline{[6.4,16.6]} & 
$^{{\scriptsize *}}$\newline{[3.0,14.6]} & 
$^{{\scriptsize -}}$\newline{[\mbox{-}0.6,11.5]} & 
$^{{\scriptsize ***}}$\newline{[1.7,3.8]} & 
$^{{\scriptsize ***}}$\newline{[1.7,4.2]} & 
$^{{\scriptsize ***}}$\newline{[3.6,6.5]} & 
$^{{\scriptsize ***}}$\newline{[4.1,7.2]} & 
$^{{\scriptsize ***}}$\newline{[4.5,7.6]} &
$^{{\scriptsize *}}$\newline{[2.7,12.3]} & 
$^{{\scriptsize *}}$\newline{[3.2,20.9]} & 
$^{{\scriptsize *}}$\newline{[3.7,19.0]} & 
$^{{\scriptsize *}}$\newline{[2.8,17.8]} & 
$^{{\scriptsize *}}$\newline{[0.2,24.8]} \\
\cmidrule{1-16}

TFF & 
$^{{\scriptsize *}}$\newline{[0.6,4.9]} & 
$^{{\scriptsize **}}$\newline{[2.8,8.3]} & 
$^{{\scriptsize -}}$\newline{[\mbox{-}1.6,4.2]} & 
$^{{\scriptsize *}}$\newline{[0.9,6.7]} & 
$^{{\scriptsize *}}$\newline{[1.2,7.4]} & 
$^{{\scriptsize ***}}$\newline{[6.0,9.6]} & 
$^{{\scriptsize ***}}$\newline{[8.4,13.2]} & 
$^{{\scriptsize *}}$\newline{[1.0,5.1]} & 
$^{{\scriptsize ***}}$\newline{[3.5,10.1]} & 
$^{{\scriptsize ***}}$\newline{[5.5,11.1]} &
$^{{\scriptsize -}}$\newline{[\mbox{-}7.1,0.4]} & 
$^{{\scriptsize -}}$\newline{[\mbox{-}1.1,14.7]} & 
$^{{\scriptsize -}}$\newline{[\mbox{-}4.3,8.6]} & 
$^{{\scriptsize *}}$\newline{[4.1,17.9]} & 
$^{{\scriptsize **}}$\newline{[15.1,27.2]} \\
\cmidrule{1-16}

EEG-Conformer & 
$^{{\scriptsize ***}}$\newline{[4.0,6.6]} & 
$^{{\scriptsize ***}}$\newline{[2.9,6.1]} & 
$^{{\scriptsize ***}}$\newline{[2.4,4.9]} & 
$^{{\scriptsize ***}}$\newline{[3.9,6.9]} & 
$^{{\scriptsize **}}$\newline{[2.1,7.2]} & 
$^{{\scriptsize ***}}$\newline{[4.0,5.5]} & 
$^{{\scriptsize ***}}$\newline{[2.6,4.2]} & 
$^{{\scriptsize ***}}$\newline{[2.9,4.7]} & 
$^{{\scriptsize ***}}$\newline{[1.8,3.5]} & 
$^{{\scriptsize ***}}$\newline{[3.0,5.1]} &
$^{{\scriptsize **}}$\newline{[9.2,18.7]} & 
$^{{\scriptsize **}}$\newline{[14.7,28.6]} & 
$^{{\scriptsize **}}$\newline{[6.1,13.9]} & 
$^{{\scriptsize *}}$\newline{[2.0,7.2]} & 
$^{{\scriptsize **}}$\newline{[2.7,8.1]} \\

\bottomrule

\label{comparison_significance}

\end{tabularx}
\end{table}

\textcolor{blue}{Figure \ref{fig:lwf_confusion_matrix} (b) presents the confusion matrix of our method on the BETA dataset. The overall prediction accuracy is lower than on the Benchmark dataset due to the BETA dataset's higher environmental noise. Between the two error patterns previously mentioned, the second error pattern occurs more frequently in the BETA dataset. Figure \ref{fig:roc} (b) shows the ROC curves on the BETA dataset. Our framework achieves the highest AUC value at 0.8408, slightly higher than EEGConformer's 0.8393, and significantly outperforming TFF's 0.8106.}

\textcolor{blue}{Figure \ref{fig:comparison_result} (c) shows the experimental results on the Nakanishi dataset. Due to the Nakanishi dataset containing fewer classes (12), the average accuracy is higher than on the Benchmark and BETA datasets. For short signal lengths (0.3s), our framework achieves 25.89\% $\pm$ 8.56\% accuracy, significantly outperforming traditional methods like CCA (13.22\% $\pm$ 3.85\%), FBCCA (13.44\% $\pm$ 2.98\%), and TRCA (18.58\% $\pm$ 5.46\%). These differences of 12.67\%, 12.45\%, and 7.31\% respectively are statistically significant ($p < 0.01$, with confidence intervals of [9.0, 16.9], [9.11, 16.06], and [2.72, 12.28]). EEGConformer achieves an accuracy of 12.22\% $\pm$ 2.93\%, significantly lower than our method ($p < 0.001$, with confidence intervals of [9.0, 16.9]), indicating that EEGConformer's feature extraction capability is limited when working with smaller datasets. TFF achieves an accuracy of 29.05\% $\pm$ 9.71\%, slightly higher than our framework, but the difference is not statistically significant. This occurs because TFF has many more model parameters than our framework, allowing it to capture deeper signal features. However, when signal length increases to 0.4s and beyond, TFF's accuracy falls below our framework, and at signal lengths of 0.5s or greater, TFF's accuracy becomes significantly lower than our framework. This demonstrates TFF's poor stability in processing EEG signals.
Figure \ref{fig:lwf_confusion_matrix} (c) shows the confusion matrix of our framework on the Nakanishi dataset. The error patterns are not as pronounced as in the Benchmark and BETA datasets because the Nakanishi dataset contains much less data than both Benchmark and BETA datasets.
Figure \ref{fig:roc} (c) presents the ROC curves for the Nakanishi dataset. Our framework and EEGConformer achieve AUC values of 0.9791 and 0.9773 respectively. Both values are high and very close, indicating that both methods achieve good classification performance on this dataset. In contrast, TFF's AUC value is only 0.7703, and its curve is not smooth, suggesting that TFF fails to identify key discriminative features in the Nakanishi dataset.}

\subsection{Model size and Inference time Analysis}
\label{sec:model_size_infer_time}
\begin{figure*}[t]
    \centering
    \includegraphics[width=0.9\linewidth]{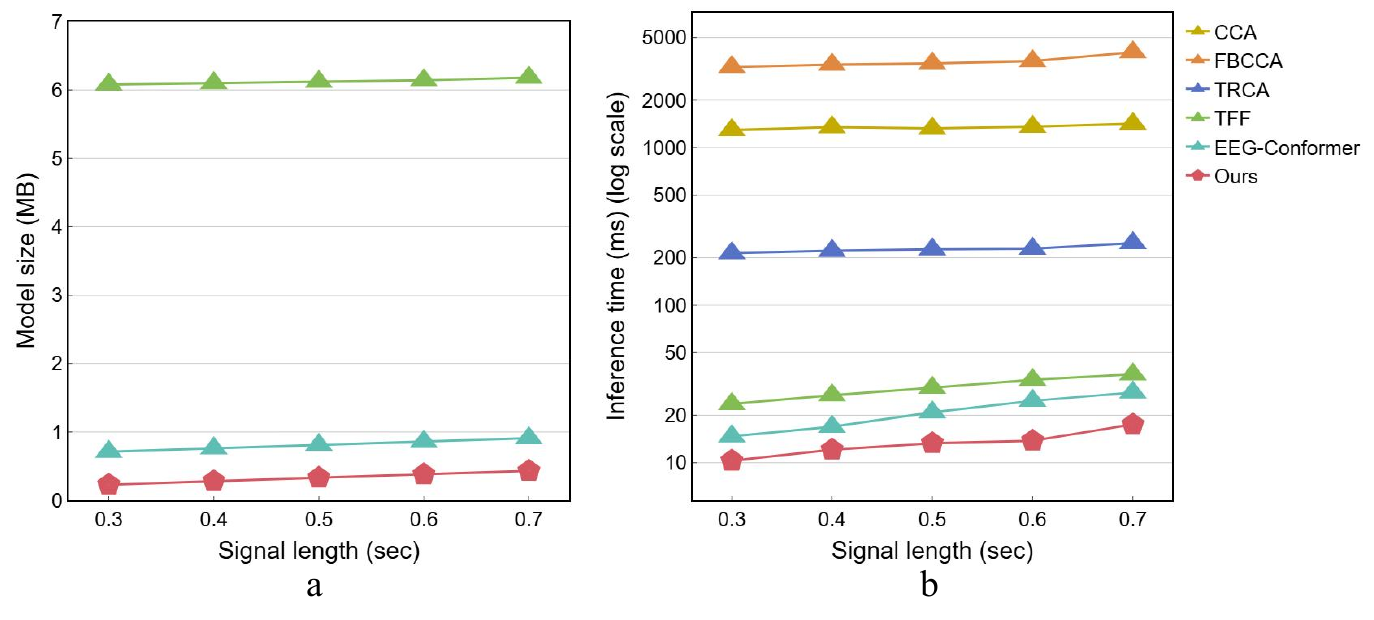}

    \caption{Comparison of Model size and Inference speed Among Different SSVEP Classification Methods. (a) Comparison of model size between our framework and baselines (b) Comparison of Inference time between our framework and baselines. 
}
    \label{fig:size_speed}
\end{figure*}

In this section, we compare our framework's model size and inference speed against other methods. We conducted these experiments using only the Benchmark dataset, as variations in data distribution do not affect measurements of model size and inference speed. Figure \ref{fig:size_speed} (a) illustrates the comparison of model sizes. As the signal length increases, the model sizes for all methods tend to grow. However, our framework consistently exhibits significantly smaller model sizes compared to TFF and EEGConformer. At a signal length of 0.7s, our framework's model size is 0.43MB, while EEGConformer and TFF require 0.91MB and 6.08MB respectively, approximately 2.1 times and 14.1 times larger than our model. When the signal length decreases to 0.3s, our framework's model size reduces to 0.23MB, while EEGConformer and TFF maintain sizes of 0.71MB and 6.07MB respectively, approximately 3.1 times and 26.4 times larger than our model. EEGConformer's larger parameter count is due to its combination of spatiotemporal convolution layers and self-attention mechanisms to capture both local and global EEG signal representations. TFF employs a transformer-based architecture with two transformer streams and thus exhibits a higher parameter count than EEGConformer. As signal length decreases, only TFF's embedding layer reduces in size, which causes minimal changes to its overall model size. Reviewing the Benchmark dataset experiments in Section \ref{sec:comparisons}, our framework outperforms TFF at signal lengths of 0.3s and 0.5s, though the differences are not significant. However, our model size analysis shows that TFF requires 26.39 and 18.39 times more parameters than our framework at 0.3s and 0.5s respectively. This demonstrates that increasing model parameters does not necessarily enhance performance. Our framework more efficiently utilizes limited trainable parameters to achieve better classification results. These experimental results demonstrate our framework achieves a significantly smaller model size. This reduction in model size makes our framework more suitable for real-world BCI system deployment.

Figure \ref{fig:size_speed} (b) presents the inference speed comparison across different methods and demonstrates that inference time increases with signal length for all approaches. Our framework demonstrates significant advantages in inference speed compared to other methods. At signal length of 0.3s, our framework completes inference in 10.3ms, while CCA and FBCCA require 1293.9ms and 3247.1ms respectively, making them 125.6 and 325.3 times slower than our framework. CCA and FBCCA require longer inference time due to their training-free nature. They need online computation of linear transformation vectors to maximize correlation between EEG signals and reference signals for classification. FBCCA exhibits longer inference time than CCA because of its additional filterbank technique, which enhances classification accuracy at the cost of inference speed. TRCA achieves an inference time of 213.8ms at signal length of 0.3s, making it 20.8 times slower than our framework. TRCA's faster processing compared to CCA and FBCCA is due to its pre-trained spatial filters. TFF and EEGConformer require 23.6ms and 14.7ms respectively, making them 2.3 and 1.43 times longer than our framework's inference time. Although TFF and EEGConformer demonstrate significant improvements over traditional methods, their inference speeds remain lower than our framework's performance. The speed difference between methods increases as signal length grows. When signal length equals 0.3s, TFF and EEGConformer are slower than our framework by 13.3ms and 4.4ms respectively. These time differences expand to 18.5ms and 10.4ms at signal length of 0.7s. This increasing gap occurs because TFF and EEGConformer use self-attention modules with $O(n^{2})$ computational complexity. 

\textcolor{blue}{Based on experimental results, our framework significantly outperforms existing methods in both model size and inference speed. This advantage is also theoretically justified for several reasons:
In the data augmentation Inter-Trial Remixing phase, our method only involves calculating means and variances, achieving a computational complexity of $O(n)$. During Context-Aware Distribution Alignment phase, we use linear layers to adjust feature distributions for each channel, with complexity only related to channel dimensions, ensuring minimal computational overhead. In the ASDM module, both the Trainable Amplitude Threshold filtering and Trainable Spectral Weighting maintain $O(n)$ complexity. This is because they perform only simple operations (amplitude comparison and element-wise multiplication respectively) once per frequency point, with computational requirements scaling linearly with signal length. While FFT and IFFT operations have $O(nlogn)$ complexity, our framework's overall computational burden remains more efficient than transformer-based approaches like EEGConformer and TFF, which have $O(n^2)$ complexity.}

Therefore, our framework demonstrates consistent speed advantages across EEG signals of different lengths. The faster inference speed of our framework enables quick SSVEP detection and immediate BCI system responses.

\subsection{Ablation Study}

\textcolor{blue}{Table \ref{Benchmark_ablation}, Table \ref{BETA_ablation} and Table \ref{Nakanishi_ablation} present the ablation study results on the Benchmark, BETA and Nakanishi datasets. }In our ablation study, we evaluated the model's performance under three conditions: the complete model, the model without the Adaptive Spectrum Denoise Module (w.o. ASDM), and the model without Data augmentation (w.o. Aug). To assess the statistical significance of the results, we employed the Wilcoxon Signed-Rank Test , as in our previous analyses. We denote significance levels using asterisks: * for $p<0.05$, ** for $p<0.01$, and *** for $p<0.001$.

For the Benchmark dataset, our complete framework outperforms the ablated frameworks across all signal lengths. Removing the ASDM significantly decreases performance, with differences of 2.7\%, 2.6\%, 2.0\%, 3.0\%, and 3.3\%, all statistically significant ($p<0.001$, $p<0.01$, $p<0.05$, $p<0.001$, $p<0.01$ respectively). This indicates that short EEG signals are more susceptible to complex environmental noise. Without ASDM, the model fails to distinguish key frequency information and focuses on important spectral components, which leads to performance degradation. The standard deviation also increases after removing ASDM (e.g., from 15.6 to 19.6 at 0.4s signal length), suggesting less stable performance due to difficulty in handling various external noise sources. Removing the data augmentation module also decreases performance, with differences of 1.2\%, 1.8\%, 1.6\%, 2.5\%, and 3.2\%. These differences are statistically significant for all signal lengths except 0.3s. The impact of data augmentation becomes more pronounced as signal length increases, particularly at 0.6s and 0.7s ($p<0.001$, $p<0.01$). This is because longer signals exhibit more signal variability, and data augmentation helps mitigate this variability, preventing the framework from overfitting to noise and improving the framework's generalization ability. The standard deviation increases across all signal lengths when the data augmentation module is removed, indicating that data augmentation consistently enhances model stability and reduces performance fluctuations.

When applied to the BETA dataset, both the complete and ablated frameworks show a decrease in performance. This decline can be attributed to the higher environmental noise and the larger subject pool in the BETA dataset (70 subjects) compared to the Benchmark dataset (35 subjects). However, the complete framework consistently outperforms its ablated versions across all signal lengths. Removing the ASDM significantly decreases performance, with differences of 2.1\%, 1.7\%, 3.1\%, 1.9\%, and 3.0\%, all statistically significant ($p<0.001$, $p<0.001$, $p<0.001$, $p<0.05$, $p<0.001$ respectively). The performance reduction is more evident for shorter signal lengths, as short EEG signals contain limited features and are more susceptible to environmental noise. This highlights ASDM's crucial role in denoising, especially in applications with short EEG signals and complex environmental noise. Removing the data augmentation module also significantly reduces performance, with differences of 2.5\%, 1.2\%, 1.4\%, 0.8\%, and 2.2\%, all statistically significant ($p<0.001$ for all). The performance decline is more substantial in the BETA dataset compared to the Benchmark dataset. This is attributed to BETA's more complex noise and larger subject pool, which emphasizes data augmentation's importance in helping the framework adapt to high EEG variability and complex environmental noise. Furthermore, the framework without data augmentation shows increased standard deviation in most signal length conditions (except for 0.7s), indicating that the data augmentation module enhances the framework's stability when processing different subjects. This conclusion aligns with the findings from the Benchmark dataset experiments.

\renewcommand{\arraystretch}{1.3} 
\begin{table}[htbp]
\centering
\fontfamily{ptm}\selectfont
\caption{ The ablation study results on the Benchmark dataset, including accuracy and standard deviation (\%). ASDM represents the Adaptive Spectrum Denoise Module, and Aug represents Data augmentation. * indicates significant differences between our complete framework and the framework with the corresponding module removed: * $p<0.05$, ** $p<0.01$, *** $p<0.001$.}

\begin{tabular}{lccccc}
\toprule
& \multicolumn{5}{c}{\textbf{Signal length(s)}} \\
\Xhline{0.5pt} 
\textbf{module selection} & \textbf{0.3s} & \textbf{0.4s} & \textbf{0.5s} & \textbf{0.6s} & \textbf{0.7s} \\
\midrule
w.o. ASDM & $26.5\pm13.6$*** & $32.8\pm19.6$** & $38.2\pm19.0$* & $42.6\pm19.1$*** & $44.8\pm18.8$** \\
w.o. Aug & $28.0\pm14.5$ & $33.6\pm16.0$* & $38.6\pm18.4$* & $42.1\pm18.8$*** & $44.9\pm18.9$** \\
\textbf{Ours} & $\boldsymbol{29.2\pm14.1}$ & $\boldsymbol{35.4\pm15.6}$ & $\boldsymbol{40.2\pm17.2}$ & $\boldsymbol{45.6\pm17.3}$ & $\boldsymbol{48.1\pm17.5}$ \\
\bottomrule
\label{Benchmark_ablation}
\end{tabular}
\end{table}

\renewcommand{\arraystretch}{1.3} 

\begin{table}[htbp]
\centering
\fontfamily{ptm}\selectfont
\caption{ The ablation study results on the BETA dataset, including accuracy and standard deviation (\%). ASDM represents the Adaptive Spectrum Denoise Module, and Aug represents Data augmentation. * indicates significant differences between our complete framework and the framework with the corresponding module removed: * $p<0.05$, ** $p<0.01$, *** $p<0.001$.}

\begin{tabular}{lcccccc}
\toprule
& \multicolumn{5}{c}{\textbf{Signal length(s)}} \\
\Xhline{0.5pt} 
\textbf{module selection} & \textbf{0.3s} & \textbf{0.4s} & \textbf{0.5s} & \textbf{0.6s} & \textbf{0.7s} \\
\midrule
w.o. ASDM & $11.5\pm6.1$*** & $16.2\pm9.4$*** & $20.3\pm12.5$*** & $25.8\pm15.0$* & $29.5\pm16.7$*** \\
w.o. Aug & $12.1\pm7.0$*** & $16.7\pm9.7$*** & $22.0\pm13.3$*** & $26.9\pm15.9$*** & $30.3\pm17.6$*** \\
\textbf{Ours} & $\boldsymbol{13.6\pm6.7}$ & $\boldsymbol{17.9\pm9.3}$ & $\boldsymbol{23.4\pm12.9}$ & $\boldsymbol{27.7\pm14.7}$ & $\boldsymbol{32.5\pm17.8}$ \\
\bottomrule
\label{BETA_ablation}
\end{tabular}
\end{table}

\renewcommand{\arraystretch}{1.3} 

\begin{table}[htbp]
\centering
\fontfamily{ptm}\selectfont
\caption{ The ablation study results on the Nakanishi dataset, including accuracy and standard deviation (\%). ASDM represents the Adaptive Spectrum Denoise Module, and Aug represents Data augmentation. * indicates significant differences between our complete framework and the framework with the corresponding module removed: * $p<0.05$, ** $p<0.01$, *** $p<0.001$.}

\begin{tabular}{lcccccc}
\toprule
& \multicolumn{5}{c}{\textbf{Signal length(s)}} \\
\Xhline{0.5pt} 
\textbf{module selection} & \textbf{0.3s} & \textbf{0.4s} & \textbf{0.5s} & \textbf{0.6s} & \textbf{0.7s} \\
\midrule
w.o. ASDM & $18.8\pm7.7$** & $33.9\pm15.1$** & $43.7\pm19.8$** & $53.8\pm21.9$** & $63.6\pm23.6$** \\
w.o. Aug & $20.5\pm7.5$** & $33.1\pm15.2$** & $43.9\pm18.4$** & $48.2\pm20.9$* & $63.1\pm25.0$* \\
\textbf{Ours} & $\boldsymbol{25.9\pm8.6}$ & $\boldsymbol{37.4\pm15.7}$ & $\boldsymbol{48.4\pm20.3}$ & $\boldsymbol{56.5\pm22.2}$ & $\boldsymbol{66.3\pm23.9}$ \\
\bottomrule
\label{Nakanishi_ablation}
\end{tabular}
\end{table}

\textcolor{blue}{When applied to the Nakanishi dataset, the complete framework demonstrates superior performance compared to its ablated versions across all signal lengths. The ablation study results in Table 4 clearly illustrate the contribution of each module to the overall framework effectiveness. Removing the ASDM significantly impairs performance, with accuracy reductions of 7.1\%, 3.5\%, 4.7\%, 2.7\%, and 2.7\% across signal lengths from 0.3s to 0.7s, all statistically significant ($p < 0.01$). This performance degradation is particularly pronounced at shorter signal lengths (0.3s), where the accuracy drops from 25.9\% to 18.8\%. This highlights ASDM's critical role in enhancing signal quality by denoising, especially when processing shorter EEG signals that contain limited informative features. Similarly, eliminating the data augmentation module results in substantial performance decreases of 5.4\%, 4.3\%, 4.5\%, 8.3\%, and 3.2\% across the five signal length conditions, all statistically significant ($p < 0.05$ or $p < 0.01$). The performance reduction is most evident at the 0.6s signal length, where accuracy falls from 56.5\% to 48.2\%. Additionally, the framework without data augmentation exhibits increased standard deviation in most conditions, indicating that data augmentation enhances the model's stability and generalization capability when processing varied subject data.}

\subsection{\textcolor{blue}{Limitation and Future work}}
\textcolor{blue}{While our framework achieved competitive performance across multiple experimental datasets, we identify a few key limitations. One limitation is that the model may overfit to specific subjects or environmental conditions, which could reduce its generalizability to other subjects or settings. Furthermore, the study relied on publicly available EEG datasets, and no real-time, real-world EEG decoding experiments were conducted, limiting the practical applicability of the approach.}

\textcolor{blue}{In future work, we aim to address the limitations identified above. Specifically, we will focus on improving the model's robustness to different subjects and varying environmental conditions to avoid overfitting. We will leverage the advantage of our lightweight model to integrate it with robotics(\cite{chen2024ddl,lee2021design,jian2024lvcp}), enabling real-time BCIs. We also plan to extend the framework to handle more diverse EEG-based tasks, including motor imagery(\cite{liu2021tacnet,zhang2021eeg,liu2022tcacnet}) and emotion analysis(\cite{ladda2021using}).}


\section{Conclusion}

In this paper, we propose a calibration-free EEG decoding framework for fast SSVEP detection. Our framework employs an adaptive data augmentation method to address the challenge of insufficient training data, enabling the model to identify vital task-relevant content effectively. Additionally, we introduce the Adaptive Spectrum Denoise Module, which removes noise from EEG signals through adaptive frequency domain analysis, thereby enhancing signal quality and model robustness. The proposed framework has been validated on three public datasets, where it outperforms both traditional methods and deep learning methods in short EEG signal classification accuracy with smaller model size and faster inference speed. Additionally, we conducted ablation studies to confirm the efficacy of our proposed module. These results demonstrate that our calibration-free framework is well-suited for fast SSVEP detection, facilitating the development and deployment of responsive and reliable brain-computer interfaces for real-world applications.

\bibliographystyle{unsrtnat} 

\bibliography{cas-refs}


\end{document}